# Producing treatment hierarchies in network meta-analysis using probabilistic models and treatment-choice criteria

**Theodoros Evrenoglou[1,2], Adriani Nikolakopoulou[1,3], Guido Schwarzer[1], Gerta Rücker[1], Anna Chaimani[2,4]**

[1]*Institute of Medical Biometry and Statistics, Faculty of Medicine and Medical Center-University of Freiburg, Freiburg im Breisgau, Germany*

[2]*Center of Research in Epidemiology and Statistics (CRESS-U1153), Université Paris Cité, INSERM, Paris, France*

[3]*Department of Hygiene, Social-Preventive Medicine and Medical Statistics, School of Medicine, Aristotle University of Thessaloniki, Thessaloniki, Greece*

[4]*Oslo Center for Biostatistics and Epidemiology, Department of Biostatistics, University of Oslo, Oslo, Norway*

Corresponding author: Dr. Theodoros Evrenoglou

Institute of Medical Biometry and Statistics, Faculty of Medicine and Medical Center - University of Freiburg, Freiburg, Germany

email: theodoros.evrenoglou@uniklinik-freiburg.de

## Abstract

A key output of network meta-analysis (NMA) is the relative ranking of treatments; nevertheless, it has attracted substantial criticism. Existing ranking methods often lack clear interpretability and fail to adequately account for uncertainty, over-emphasizing small differences in treatment effects. We propose a novel framework to estimate treatment hierarchies in NMA using a probabilistic model, focusing on a clinically relevant treatment-choice criterion (TCC). Initially, we formulate a mathematical expression to define a TCC based on smallest worthwhile differences (SWD), converting NMA relative treatment effects into treatment preference format. This data is then synthesized using a probabilistic ranking model, assigning each treatment a latent "ability" parameter, representing its propensity to yield clinically important and beneficial true treatment effects relative to the rest of the treatments in the network. Parameter estimation relies on the maximum likelihood theory, with standard errors derived asymptotically from Fisher's information matrix. To facilitate the use of our methods, we launched the R package **mtrank**. We applied our method to two clinical datasets: one comparing 18 antidepressants for major depression and another comparing 6 antihypertensives for the incidence of diabetes. Our approach provided robust, interpretable treatment hierarchies that account for a concrete TCC. We further examined the agreement between the proposed method and existing ranking metrics in 153 published networks, concluding that the degree of agreement depends on the precision of the NMA estimates. Our framework offers a valuable alternative for NMA treatment ranking, mitigating over-interpretation of minor differences. This enables more reliable and clinically meaningful treatment hierarchies.

**Keywords:** clinically important difference, network of interventions, probabilistic regression model, smallest worthwhile difference, treatment hierarchy

# 1 Introduction

Interpretation of network meta-analysis (NMA) outputs can be challenging as it usually comprises consideration of multiple treatment effects with different levels of uncertainty and credibility across comparisons in the network[1,2]. For example, in the relatively simple case of a network with 6 treatments the output of NMA consists of 15 treatment effect estimates. In such a context, treatment ranking can be a reliable way to summarize the evidence provided by complex treatment networks[1,3,4]. This may explain the fact that treatment hierarchies are frequently presented in published NMAs with 43% of them reporting at least one ranking metric[5].

Probably the most commonly used ranking metric, until recently, was the probability of a treatment to have the best value[5], usually denoted as $p_{BV}$. This is primarily a Bayesian metric, but it can also be calculated within the frequentist framework using resampling, thereby mimicking a Bayesian framework with flat priors. It represents the probability that a treatment in the network will have the best true treatment effect[6]. Although $p_{BV}$ has been widely used in published NMAs, more recently it has been criticized for not properly accounting for the uncertainty of the NMA estimates[4,7,8].

Other common ranking metrics are P-scores[4], which are obtained analytically through the cumulative density function of the standard normal distribution, or their Bayesian equivalent SUCRA[1] that represent the surface under the cumulative ranking curve for each treatment. The main limitation of these metrics is that they often lead to attributing distinct ranks to treatments even when there are only small differences between their SUCRA values or P-scores. Nikolakopoulou et al.[7] employed the "deviation from the means" approach for the construction of the design matrix in the NMA model and introduced a new ranking metric, called the probability of a treatment being preferable to a fictional treatment of average performance (PReTA). This metric potentially accounts better for the uncertainty in the relative effects than P-scores or SUCRAs, particularly when there is substantial variability in the precision of the NMA estimates. This is an important advantage since an empirical study revealed high agreement across all ranking metrics when NMA estimates had similar variance estimates, but large sensitivity to the choice of metric for networks with large discrepancies in the variance of the NMA estimates[5]. More recently, new ranking metrics and approaches have been developed to address more complex ranking questions. Mavridis et al.[9] extended P-scores to incorporate clinically important values, while Curteis et al. proposed a similar extension in terms of the SUCRA ranking method[10]. Chaimani et al.[11] suggested that treatment rankings should consider not only the summary relative effects but also other information, such as study or treatment characteristics. They introduced a new metric, called the probability of selecting a treatment to recommend (POST-R) that implements additional characteristics in treatment hierarchy (e.g. risk of bias or treatment cost). Papakonstantinou et al.[12] developed a resampling approach for estimating the probability that a specific treatment hierarchy occurs or a predefined criterion may be met.

Despite its usefulness when properly reported and interpreted, treatment ranking in NMA has been accompanied with a lot of skepticism[13–16]. Other common arguments against treatment ranking include that it can be biased, it is difficult to interpret, it is not accompanied with uncertainty measures, and it may overemphasize non-important differences in the treatment effect estimates[14,15]. For example, Kibret et al. performed a simulation study and found that ranking can be biased when there is an unequal number of studies per comparison in the network, with the rank probability for the treatment included in the fewest number of studies tending to suffer from upward bias[16]. However, Salanti et al.[6] argued that these criticisms should not refer to the ranking metrics per se but to the way they are used and interpreted. This is because different metrics target different types of hierarchy questions and researchers should clearly define what they mean by "best treatment" in a given setting. Hence, setting a well-defined treatment hierarchy question should always precede the estimation of treatment ranking and drive the choice of the ranking metric[6].

In this article, we introduce a novel approach for estimating treatment hierarchies in NMA based on a treatment choice criterion (TCC) constructed to ensure clinically important treatment effects. This TCC splits the NMA estimates into those fulfilling the criterion - indicating a treatment effect justifying a treatment preference - and those without a clear treatment preference. We then use a probabilistic model that yields the final treatment hierarchy by synthesizing the treatment preferences obtained from the TCC. Our manuscript is organized as follows. First, we formulate a mathematical expression that defines the TCC based on clinically important values. We then apply the criterion to the NMA treatment effects, taking into account their confidence intervals to get either a treatment preference or a tie. Our synthesis model estimates the treatment hierarchy through a latent parameter assigned to each treatment in the network that represents its 'ability' to yield clinically important and beneficial treatment effects in context of the defined TCC. In this way, treatments with higher estimated abilities are positioned more prominently in the final ranking. This modeling approach has been previously used to produce rankings in fields outside of medicine, such as sports science[17], animal behavior[18] and risk analysis[19]. To illustrate our method and compare it with existing alternatives we use two published NMAs: one comparing different antidepressants[20] for major depression and a second evaluating different antihypertensives[21] for the incidence of diabetes. Finally, we investigate the agreement between the new and existing ranking metrics through an empirical study where we re-analyse 153 published networks[22,23].

# 2 Methods

## 2.1 Defining treatment choice criteria based on NMA estimates

Suppose a network of $N$ studies comparing $T$ treatments. Let $\widehat{\boldsymbol{\theta}} = \left[\hat{\theta}_{XY}\right]_{X \neq Y}$, where $X, Y \in \{1,2,\dots,T\}$, denote the $\binom{T}{2}$-vector containing all treatment effect estimates obtained from the NMA. Let also $\boldsymbol{l} = [l_{XY}]_{X \neq Y}$ and $\boldsymbol{u} = [u_{XY}]_{X \neq Y}, X, Y \in \{1,2,\dots,T\}$ represent the corresponding vectors containing the lower and upper bounds of the confidence intervals for each $\hat{\theta}_{XY}$. We start building our modeling approach by defining concrete criteria for choosing one treatment over another or considering two treatments as equivalent. These criteria have the form of a decision rule and may depend on several factors, such as the clinical setting, the outcome(s) under investigation, or even the type of patients under consideration (e.g. chronic patients vs treatment-naïve individuals). Here, we suggest a generic approach that can be easily adapted to different settings based on the so-called range of equivalence (ROE). The ROE has been previously introduced as a way to infer on the clinical importance of a treatment effect in the context of appraising NMA estimates; relative effects lying within this range are considered lacking a treatment preference[24].

Following Nikolakopoulou et al.[24], we construct the ROE using the smallest worthwhile difference (SWD) and its reciprocal (or opposite) value. The SWD represents the smallest beneficial effect of a treatment that justifies a preference for it over another treatment[25]. A comparison between treatments $X$ and $Y$ fulfils the TCC and indicates a treatment preference if the relative treatment effect $\hat{\theta}_{XY}$ and one or both of the confidence interval bounds are outside the ROE. Then, a treatment preference (i.e., $Y = X$ or $X > Y$) is determined based on the direction of the treatment effect, favoring either $X$ or $Y$, and the bounds of its confidence interval, which should indicate a clinically important treatment effect. Otherwise, in the absence of evidence to support a clear treatment preference, the TCC is not satisfied and therefore the treatments are considered as equivalent (i.e. $X = Y$). To mathematically represent this rule, for the case of a beneficial outcome (i.e. larger treatment effect values are desirable), suppose that $\mathrm{I}_{XY}^{(1)}$ and $\mathrm{I}_{XY}^{(2)}$ are two indicator variables defined for each pairwise comparison $XY$ in the network with $X, Y = 1,2, \dots, T,$ as

$$\mathrm{I}_{XY}^{(1)} = \begin{cases} 1, \text{if } \left(\hat{\theta}_{XY} > U^{\mathrm{ROE}}\right) \text{ and } \left(l_{XY} > L^{ROE}\right) \\ 0, \text{otherwise} \end{cases} \quad (1)$$

and

$$\mathrm{I}_{XY}^{(2)} = \begin{cases} -1, \text{if } \left(\hat{\theta}_{XY} < L^{\mathrm{ROE}}\right) \text{ and } \left(u_{XY} < U^{ROE}\right) \\ 0, \text{otherwise} \end{cases} \quad (2)$$

where $L^{\mathrm{ROE}}$ and $U^{\mathrm{ROE}}$ are the lower and upper limits of the ROE respectively. In this way the indicator variable $\mathrm{I}_{XY}^{(1)}$ becomes 1 when the necessary TCC conditions for the treatment $Y$ to be preferred over treatment $X$ are satisfied and 0 elsewhere. Similarly, the indicator variable $\mathrm{I}_{XY}^{(2)}$ becomes -1 when the necessary TCC conditions for treatment $X$ to be preferred over the treatment $Y$ are satisfied and 0 elsewhere. Then, the TCC for each comparison $XY$ can be defined for a beneficial outcome based on the following conditions:

$$Y > X, if \sum_{k=1}^{2} \mathrm{I}_{XY}^{(\mathrm{k})} = 1 \quad (3a)$$

$$X > Y, \ if \sum_{k=1}^{2} \mathrm{I}_{XY}^{(\mathrm{k})} = -1 \quad (3b)$$

$$Y = X, \ if \sum_{k=1}^{2} \mathrm{I}_{XY}^{(\mathrm{k})} = 0 \quad (3c)$$

where $Y > X$ or $X > Y$ indicate a treatment preference and $Y = X$ indicates a 'tie', representing cases where the available evidence from the NMA estimate for the comparison $XY$ does not support any clear treatment preference, based on the TCC. A graphical illustration of the above TCC for the case of a beneficial outcome in a fictional example is given in **Figure 1**. In case of a harmful outcome, we need to reverse the signs of 1 and -1 in Equations (3a) and (3b). Throughout the rest of the manuscript, treatment effects that justify a treatment preference according to the TCC will also be referred to as clinically important effects. Approaches for defining a SWD have been suggested elsewhere and are beyond the scope of this article[25–27]. Investigators who prefer to use a different TCC can modify Equations (1) and *(2)* accordingly. After defining the TCC, all the $\binom{T}{2}$ NMA estimates are transformed into a treatment preference format using Equations (3a)-(3c) for each comparison $XY$ in the network.

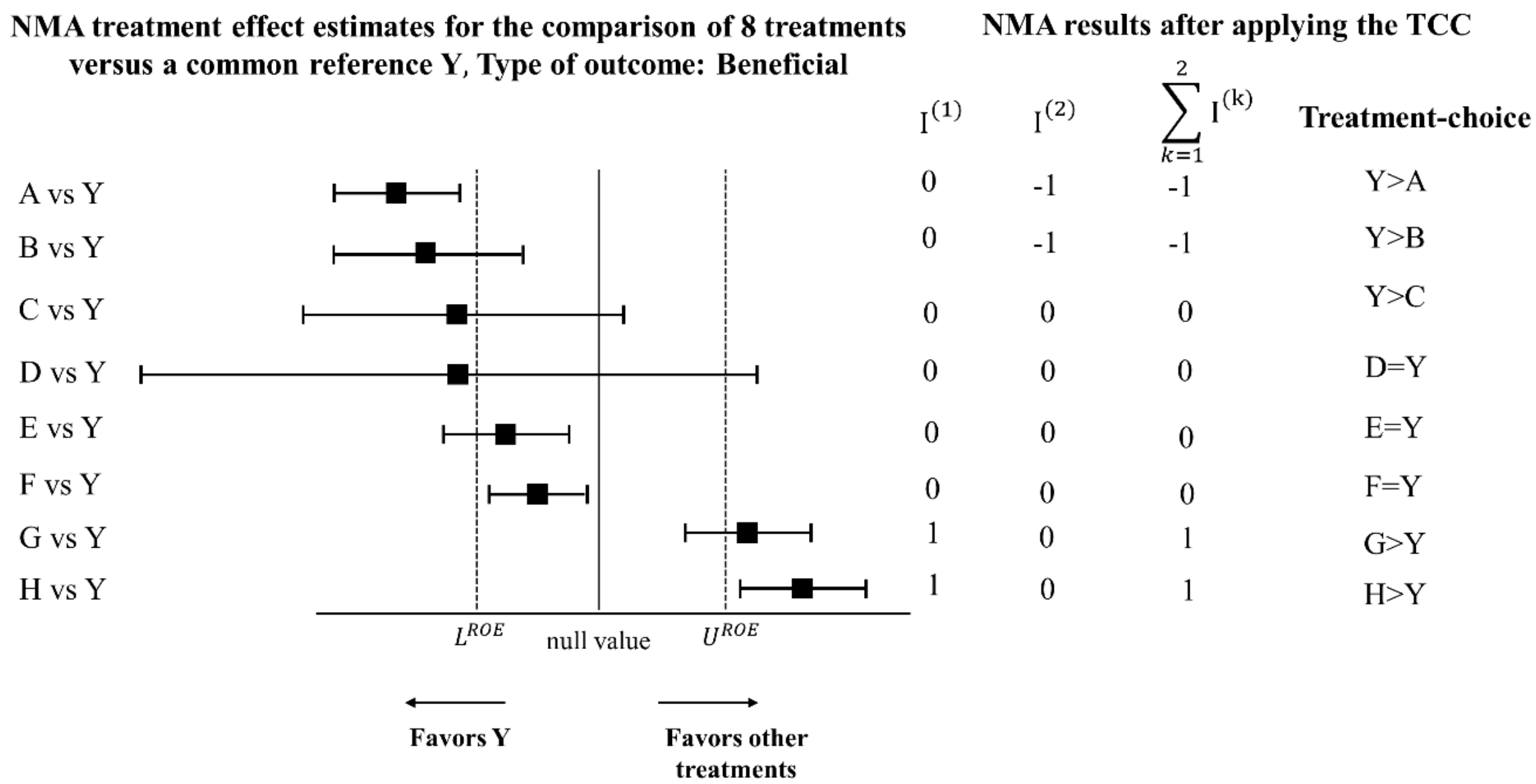

**Figure 1:** A graphical representation of the TCC for a fictional example showing the NMA estimates for the comparison of 8 treatments versus a common reference treatment Y in terms of a beneficial outcome.

## 2.2 Estimating treatment hierarchies using probabilistic models

To synthesize the resulting treatment preference data, we adapt the so-called 'Bradley-Terry model'[28–31] to the context of NMA. This is a probabilistic model, suitable for modeling preference data and originally suggested to estimate ranking outside NMA (e.g. sports tournaments) but, to our knowledge, was never adapted to estimate treatment hierarchies in NMA. We parameterize the model using an unobserved latent parameter $\psi_X \geq 0$ that represents the 'ability' each treatment $X$ has to outperform the other treatments in the network given the TCC. Here, the term 'ability to outperform' refers to the propensity of a treatment $X$ to yield clinically important and beneficial true treatment effects when compared to the rest of the treatments in the network. For the remainder of the manuscript, the term 'outperform' will be used according to this definition. Given all these considerations, the treatment hierarchy question addressed here is "Based on the pre-defined TCC, what is the ability of each treatment to yield clinically important and beneficial true treatment effects when compared to the rest of the treatments in the network?" Consequently, the ability of each treatment to fulfil the TCC is the latent parameter of interest and is modelled through a Bradley-Terry model[28–31].

Establishing a direct one-to-one mathematical relationship between the true treatment effects and the treatment abilities is challenging as the former are fixed relative unknown parameters, while the latter are unobserved latent treatment characteristics specific to the TCC of interest. Nevertheless, it is expected that treatments associated with estimated treatment effects fulfilling the TCC will yield larger ability estimates. In this way, treatments with higher estimated ability occupy a higher position in the final ranking list[29]. Let also $\boldsymbol{\psi}$ denote the $T$-length vector that contains the ability of each treatment in the network. The idea behind this model stems from Luce's axiom[29,32,33] of choice which states that the probability that a treatment $X$ has the largest ability among all $T$ treatments, with respect to the TCC, is equal to $\frac{\psi_X}{\sum_{i=1}^{T}\psi_i}$.

### 2.2.1 Synthesizing treatment preferences in the absence of ties

Following the above axiom, for each pairwise comparison in the network the probability that treatment $X$ will outperform treatment $Y$ ($X \neq Y;\ X, Y, = 1,2, \dots, T$) is,

$$\Pr(X > Y) = \frac{\psi_X}{\psi_X + \psi_Y} \tag{3}$$

with $\psi_X \geq 0\ \forall X \in \{1,2, \dots, T\}$ and $\sum_{i=1}^{T}\psi_i = 1$. Based on Equation (3) a logit-linear model can be parametrized as

$$\text{logit}(\Pr(X > Y)) = \log(\psi_X) - \log(\psi_Y) \tag{4}$$

In the present manuscript the estimation process for the above model refers to the frequentist framework and relies on maximum likelihood theory[28,34]. Fitting the model in the Bayesian setting is also possible and has been discussed elsewhere[35,36]. Let $r_{XY}$ denote a variable that takes the value 1 if , based on the TCC, treatment $X$ is preferred over treatment $Y$ and 0 otherwise. Then, the log-likelihood function across all available comparisons in the network is

$$\text{L}(\boldsymbol{\psi}) = \sum\sum_{X \neq Y} r_{XY} log\left(\frac{\psi_X}{\psi_X + \psi_Y}\right) + r_{YX} log\left(\frac{\psi_Y}{\psi_X + \psi_Y}\right), \tag{5}$$

with $\sum_{i=1}^{T}\psi_i = 1$. Maximizing the binomial log-likelihood in Equation (5) results in the maximum likelihood estimates (MLE) $\hat{\psi}_X, \forall X \in \{1,2, \dots, T\}$ of the ability parameters.

Equation (3) assumes that one treatment is always preferred over another for any pairwise comparison in the network. However, this can violate the TCC defined in Section 2.1 where we also consider that two treatments may not justify a treatment preference. Therefore, an extension to this ranking model is necessary to allow incorporating ties between treatments.

### 2.2.2 Synthesizing treatment preferences including ties

Following Davidson[31], here we assume that the probability of a tie between two treatments $X$ and Y relates to $\nu\sqrt{\psi_X\psi_Y}$. The quantity $\sqrt{\psi_X\psi_Y}$ is the geometric mean of $\psi_X$ and $\psi_Y$, while $\nu$ is a scalar nuisance parameter that describes the prevalence of ties in the network. Hence, the probability that $X$ outperforms $Y$ is now

$$\Pr(X > Y) = \frac{\psi_X}{\psi_X + \psi_Y + \nu\sqrt{\psi_X\psi_Y}} \tag{6}$$

the probability $Y$ outperforms $X$ is

$$\Pr(Y > X) = \frac{\psi_Y}{\psi_X + \psi_Y + \nu\sqrt{\psi_X\psi_Y}} \tag{7}$$

and the probability that the two treatments are tied is

$$\Pr(X = Y) = \frac{\nu\sqrt{\psi_X\psi_Y}}{\psi_X + \psi_Y + \nu\sqrt{\psi_X\psi_Y}} \tag{8}$$

with $\psi_X \geq 0, \forall X \in \{1,2, \dots T\}, \nu > 0$ and $\sum_{i=1}^{T} \psi_i = 1$. Note that parametrizing the probability of a tie using Equation (8) offers the mathematical convenience that, for a fixed value of $\nu$, the probability of a tie is maximized when $\psi_X = \psi_Y$. In other words, the probability of a tie is maximized between treatments with equal abilities. The mathematical proof of this is provided in the Supplementary Material. Considering Equations (6)-(8), the log-likelihood in Equation (5) becomes,

$$L^*(\boldsymbol{\psi}, \nu) = \sum \sum_{X \neq Y} r_{XY} \log \left( \frac{\psi_X}{\psi_X + \psi_Y + \nu\sqrt{\psi_X \psi_Y}} \right) + r_{YX} \log \left( \frac{\psi_Y}{\psi_X + \psi_Y + \nu\sqrt{\psi_X \psi_Y}} \right) + w_{XY} \log \left( \frac{\nu\sqrt{\psi_X \psi_Y}}{\psi_X + \psi_Y + \nu\sqrt{\psi_X \psi_Y}} \right) \tag{9}$$

with $\sum_{i=1}^{T} \psi_i = 1$ and $\nu > 0$. Here $w_{XY}$ is the tie variable that takes the value 1 if the TCC indicates that $X = Y$; otherwise it is equal to 0. Maximization of the multinomial log-likelihood in Equation (9) relies on iterative optimization processes such as the Newton-Raphson[31] or the minorization-maximization[34] algorithms and results in the MLEs of the ability parameters $\boldsymbol{\psi}$ and the scalar parameter $\nu$ that represents the prevalence of ties.

The asymptotic distribution of $\widehat{\boldsymbol{\psi}}$ is a multivariate normal distribution with mean $\boldsymbol{\psi}$ and variance-covariance matrix $\boldsymbol{\Sigma}^{-1}$ obtained as the inverse of the Hessian matrix $\boldsymbol{\Sigma}$. The elements of $\boldsymbol{\Sigma}$ correspond to the second partial derivatives of the log-likelihood in Equation (9). Based on the asymptotic theory, the standard errors of the elements of $\widehat{\boldsymbol{\psi}}$ are derived as the square roots of the diagonal elements of matrix $\boldsymbol{\Sigma}^{-1}$. Finally, a unique and positive MLE for each $\psi_X, X \in \{1,2., \dots, T\}$ exists under Ford's regularity condition[37]. This requires that for every possible partition of the treatments into two non-empty subsets, some treatments in the second subset are preferred to some treatments in the first subset at least once. In other words, according to the proposed methodology, if only ties are obtained from the TCC, it is not meaningful to estimate any treatment hierarchy.

### 2.2.3 Absolute and relative treatment abilities

Maximizing Equation (9) in terms of $\boldsymbol{\psi}$ refers to an optimization problem constrained at the region $\{\psi_X \geq 0,\ \sum_{i=1}^{T} \psi_i = 1\}$. This constraint prevents from negative estimates of the ability parameters and guarantees that the optimization problem remains identifiable. Then, the resulting $\widehat{\psi}_X$ represents the estimated absolute abilities of each treatment in the network. However, as also noted elsewhere[29], the scale of the absolute ability estimates is immaterial; what matters here is the relative comparison between abilities. To address this issue, we construct an artificial reference treatment group[30] $T + 1$, with ability equal to the average of the absolute ability estimates across all the $T$ treatments. This implies that we assume the ability of the treatment $T + 1$ being equal to $\psi_{T+1} = \frac{\sum_{i=1}^{T} \hat{\psi}_i}{T}$. Then, the ranking results are presented in terms of the ability ratios $\frac{\hat{\psi}_X}{\psi_{T+1}} \forall X \in \{1,2, \dots, T\}$.

The final estimates $\hat{\psi}_X$ do not necessarily satisfy $\sum_{i=1}^{T} \psi_i = 1$ as the re-normalization of the vector $\boldsymbol{\psi}$ is not needed after each iteration of the iterative process[34]. However, based on Luce's axiom of choice[32,33], we can re-normalize the absolute ability estimates as $\hat{\pi}_X = \frac{\hat{\psi}_X}{\sum_{i=1}^{T} \hat{\psi}_i}$. This allows interpreting $\hat{\pi}_X$ as the probability that each treatment $X \in \{1,2, \dots, T\}$ has the largest true ability to yield clinically important and beneficial treatment effects, with respect to the TCC, among all the $T$ treatments in the network. This additional probabilistic ranking metric, $\hat{\pi}_X$, offers a straightforward interpretation but it does not account for the uncertainty of the ability estimates $\hat{\psi}_X$. Therefore we propose $\hat{\pi}_X$ be presented alongside the ability estimates $\hat{\psi}_X$, particularly when these estimates are derived with similar levels of uncertainty in the top positions of the ranking list.

# 3 Applications

We illustrate the use of our treatment ranking method and compare it with existing ranking approaches using two published networks. The first compares the efficacy of several antidepressants for major depression[20] and the second compares different antihypertensive treatment classes and placebo for the incidence of diabetes[21]. We compared five ranking approaches: (a) P-scores[4], (b) P-scores 'adjusted' for the SWD[9], (c) the PReTA-ranking[7], (d) the ranking according to $p_{BV}$ in the frequentist setting, and (5) the estimated treatment abilities from our ranking approach. All ranking metrics were calculated based on a random-effects NMA model. To conduct the analysis, we used R version 4.4.1 (2024-06-14) and we used the R package **netmeta**[38] to fit the NMA models. To facilitate the use of our proposed approach we have created the R package **mtrank**[39] which is available on CRAN.

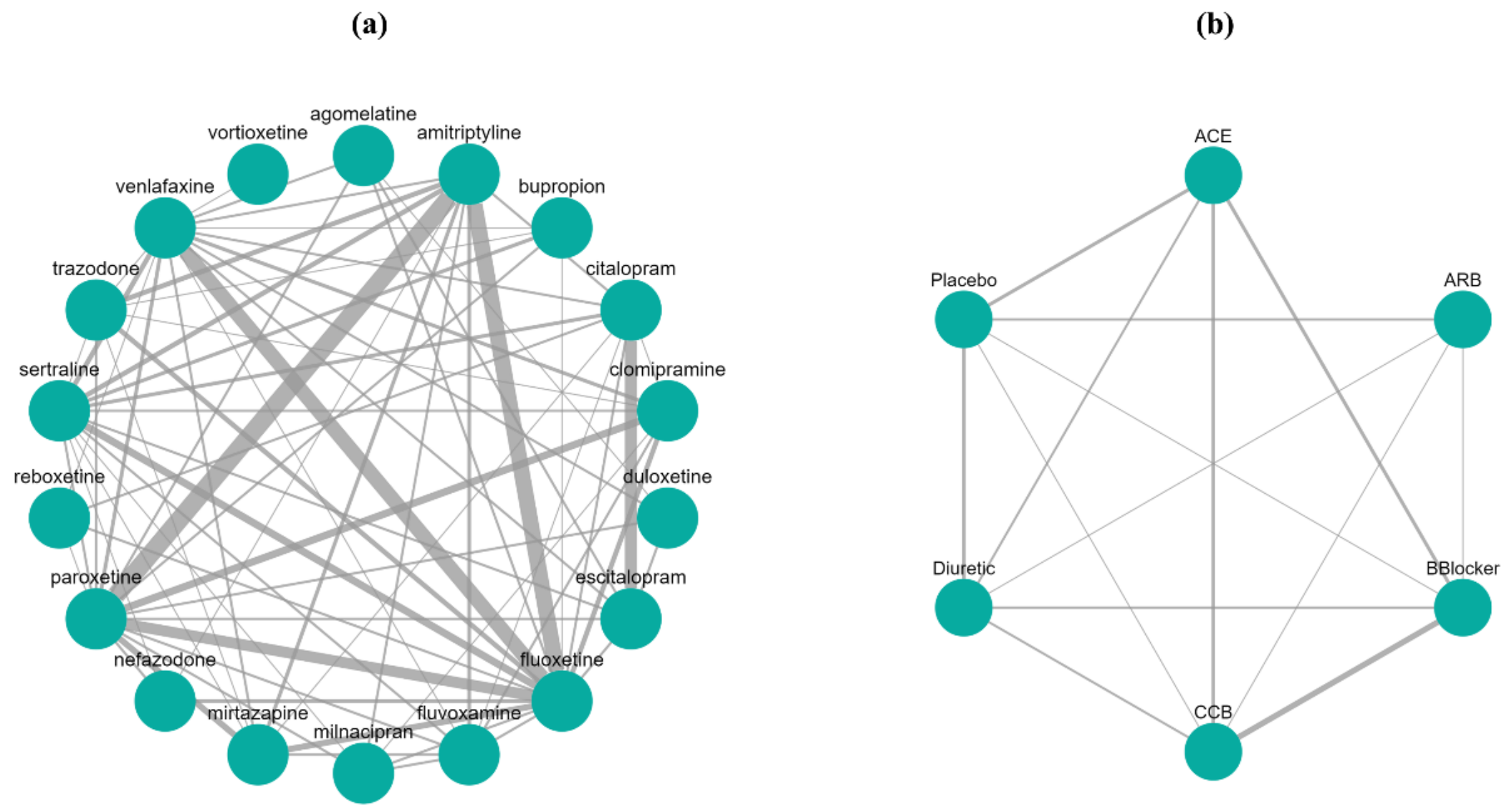

**Figure 2:** Network plots for the two clinical examples. Panel (a) shows the network of antidepressants while panel (b) the network of antihypertensive treatments. Abbreviations for panel (b), ACE: angiotensin-converting-enzyme inhibitors, ARB: angiotensin-receptor blockers, CCB: calcium-channel blocker, BBlocker: Beta Blocker.

## 3.1 Antidepressants for major depression

This network comprises 179 trials comparing 18 antidepressant drugs (**Figure 2a**). The primary outcome is response to treatment defined as a 50% or greater reduction in a depression symptom scale between baseline and eight weeks of follow-up. The outcome is measured as odds ratios (OR).

The results for methods (a) to (d) are presented in **Table 1**, alongside the respective NMA estimates of all treatments versus Trazodone. In this network, large treatment effect values indicate beneficial effects. A consensus is observed in terms of the best treatment for the P-scores and the $p_{BV}$ which rank Vortioxetine first, while using the PReTA-ranking Escitalopram is placed at the first position and Vortioxetine second. Results in terms of median ranks are available in Supplementary Material, Table 1. These results show that Vortioxetine, Escitalopram, and Bupropion occupy the top three positions, although there is considerable uncertainty. The median ranks and 95% CIs were 1 [1, 15], 3 [1, 10], and 3 [1, 15], respectively.

**Table 1:** Ranking metrics for the network of antidepressants. Treatments with the top three values for each respective metric are shown in bold. The "Treatment" column is ordered according to P-scores.

| **Treatment** | **Odds ratio of Treatment vs Placebo [position in ranking]** | **P-scores** | **P-scores (SWD)** | **PReTA** | $p_{BV}$ | $\hat{\pi}_X$ |
|---|---|---|---|---|---|---|
| Vortioxetine | **1.87 [1]** | **0.90** | **0.75** | **0.93** | **0.64** | **0.20** |
| Escitalopram | **1.51 [3]** | **0.83** | **0.49** | **0.98** | **0.08** | **0.36** |
| Bupropion | **1.55 [2]** | **0.79** | **0.53** | 0.87 | **0.19** | **0.20** |
| Mirtazapine | 1.44 [4] | 0.75 | 0.39 | **0.91** | 0.03 | 0.06 |
| Amitriptyline | 1.40 [5] | 0.71 | 0.33 | 0.88 | 0.01 | 0.06 |
| Agomelatine | 1.36 [6] | 0.64 | 0.29 | 0.74 | 0.01 | 0.06 |
| Paroxetine | 1.34 [7] | 0.62 | 0.25 | 0.83 | 0.00 | 0.01 |
| Venlafaxine | 1.34 [8] | 0.61 | 0.25 | 0.78 | 0.00 | 0.01 |
| Duloxetine | 1.29 [9] | 0.52 | 0.21 | 0.53 | 0.01 | 0.01 |
| Milnacipran | 1.27 [10] | 0.49 | 0.19 | 0.46 | 0.01 | 0.01 |
| Sertraline | 1.25 [11] | 0.45 | 0.15 | 0.38 | 0.00 | 0.00 |
| Nefazodone | 1.18 [13] | 0.38 | 0.16 | 0.33 | 0.02 | 0.00 |
| Citalopram | 1.20 [12] | 0.37 | 0.12 | 0.24 | 0.00 | 0.00 |
| Clomipramine | 1.13 [15] | 0.26 | 0.07 | 0.10 | 0.00 | 0.00 |
| Fluvoxamine | 1.12 [16] | 0.25 | 0.07 | 0.10 | 0.00 | 0.00 |
| Fluoxetine | 1.13 [14] | 0.23 | 0.06 | 0.01 | 0.00 | 0.02 |
| Trazodone | 1 [17] | 0.12 | 0.03 | 0.02 | 0.00 | 0.01 |
| Reboxetine | 0.95 [18] | 0.09 | 0.02 | 0.02 | 0.00 | 0.01 |

The NMA treatment effect estimates for the comparison of each treatment versus Trazodone are also shown in **Figure 3a**. Overall, all NMA treatment effect estimates favor the other treatments over Trazodone. Vortioxetine has the largest treatment effect and ranks first, but it also has the largest standard error. When using $p_{BV}$, the ranking does not fully account for the uncertainty in treatment effect estimates. This explains why Vortioxetine appears to be clearly the best treatment according to $p_{BV}$.

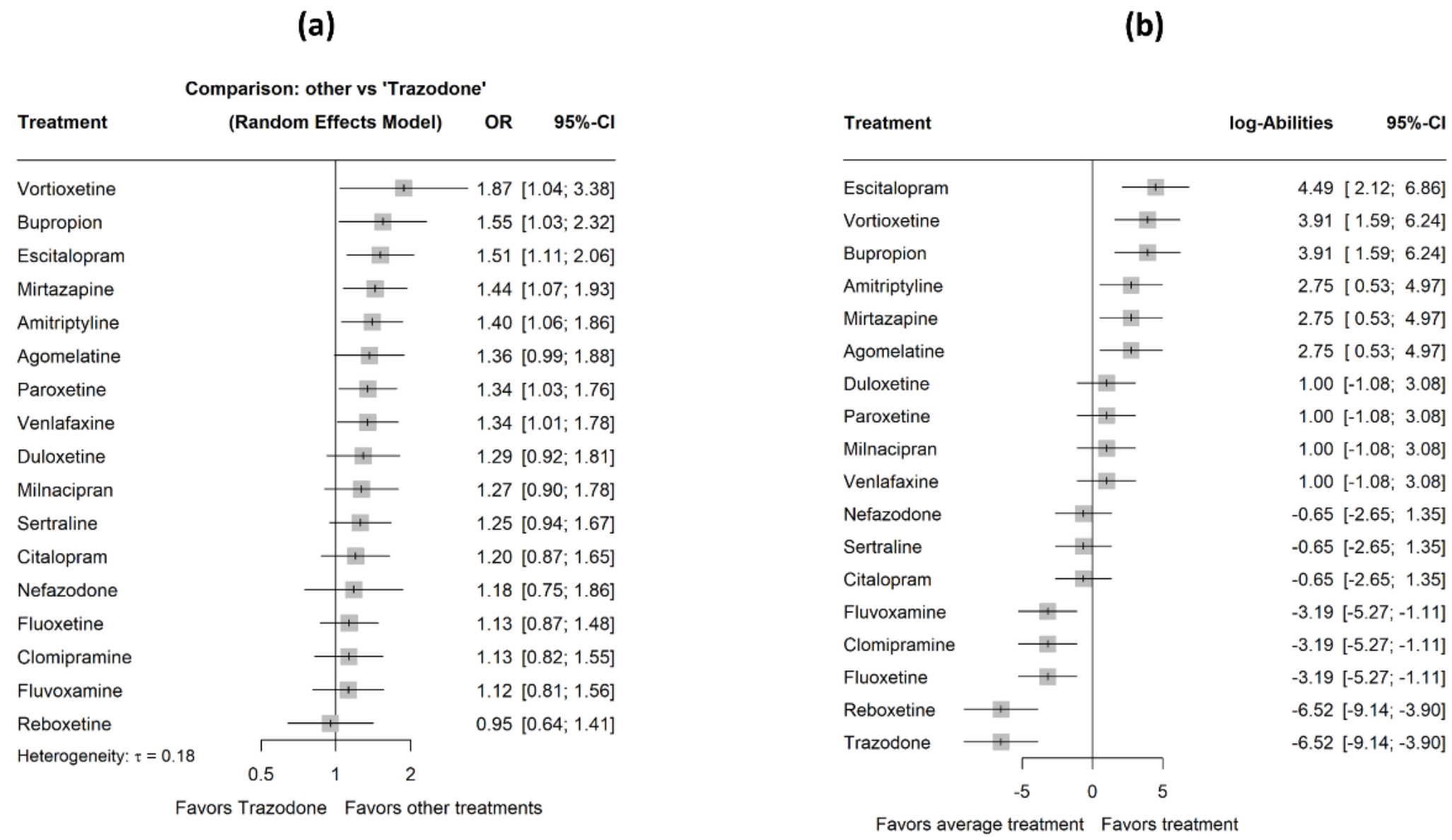

**Figure 3:** Forest plots with results for the network of antidepressants. Panel (a) shows the summary odds ratios obtained assuming Trazodone as the reference treatment group. Panel (b) shows the ranking results obtained using the proposed methodology.

Following the original publication[20], we assume a SWD equal to 1.20. Using SWD adjusted P-scores, Vortioxetine was ranked at the top position and clearly higher than Bupropion which is at the second position. The differences between unadjusted and SWD adjusted P-scores can be attributed to the increased emphasis that the latter approach puts on the magnitude of the NMA estimates. Note that the adjusted P-scores approach affects only the numerical values of the unadjusted P-scores and is generally not expected to alter the treatment hierarchy. Overall, the differences across the different hierarchies may be explained by the substantial variation of the standard errors across the NMA estimates that range from 0.07 to 0.33. The full distribution of the standard errors across all NMA estimates is depicted in Supplementary Material, Figure 1.

Setting again an SWD of 1.20, we obtain the respective ROE that ranges from 0.83 to 1.20. Then, we applied the TCC of **Section 2.1** to transform the 153 NMA estimates into treatment preferences. A high prevalence of ties was observed in the network, as only 32% of all comparisons yielded clinically important NMA estimates according to the defined TCC. The ability estimates are shown in **Figure 3b** while the normalized ability estimates $\hat{\pi}_X$ are shown in **Table 1.** Overall, within the context of the pre-defined TCC, Escitalopram demonstrated the highest ability to fulfil the TCC and yield beneficial treatment effect estimates, followed by Vortioxetine and Bupropion which are tied at the second position. In addition to these three treatments, Amitriptyline, Mirtazapine, and Agomelatine were also found to have significantly greater abilities to yield clinically important estimates than the average treatment in the network. Finally, we conducted a sensitivity analysis regarding the definition of the SWD, progressively increasing it by 0.10 increments from the value of 1.10 up to 1.50. The results are shown in **Figure 4** where to improve visibility we presented the results only in terms of the first six treatments, as per the primary analysis. Overall, the sensitivity analysis indicated that if smaller treatment effects are of interest (i.e., SWD≤1.20), then Escitalopram outperforms the other treatments. However, as the SWD increases—meaning that larger treatment effects are of interest—Vortioxetine demonstrates a greater ability to yield clinically important treatment effect estimates compared to all other treatments in the network.

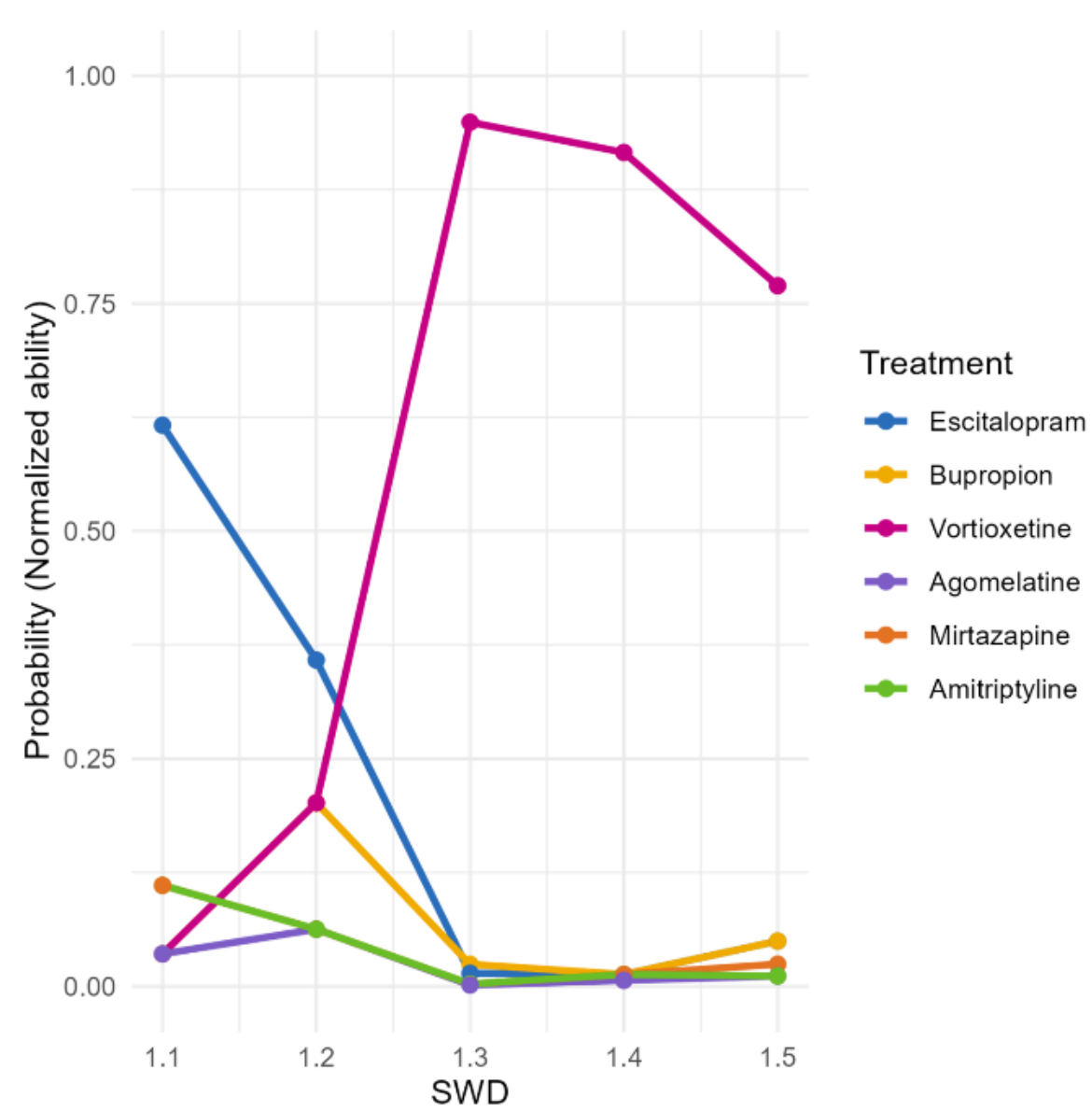

**Figure 4:** Sensitivity analysis for the network of antidepressants. The y-axis represents the probability of each treatment having the highest true ability and the x-axis the different SWD values.

## 3.2 Antihypertensive treatments and the incident of diabetes

This network consists of 22 trials comparing 5 classes of antihypertensive treatments and placebo for the incidence of diabetes[21]. This is a very well-connected network with 14 out of the 15 possible direct comparisons being observed (**Figure 2b**). The primary outcome is the proportion of patients who developed diabetes and the NMA estimates using placebo as reference can be found in **Figure 5a**. The outcome is again measured as odds ratios (OR).

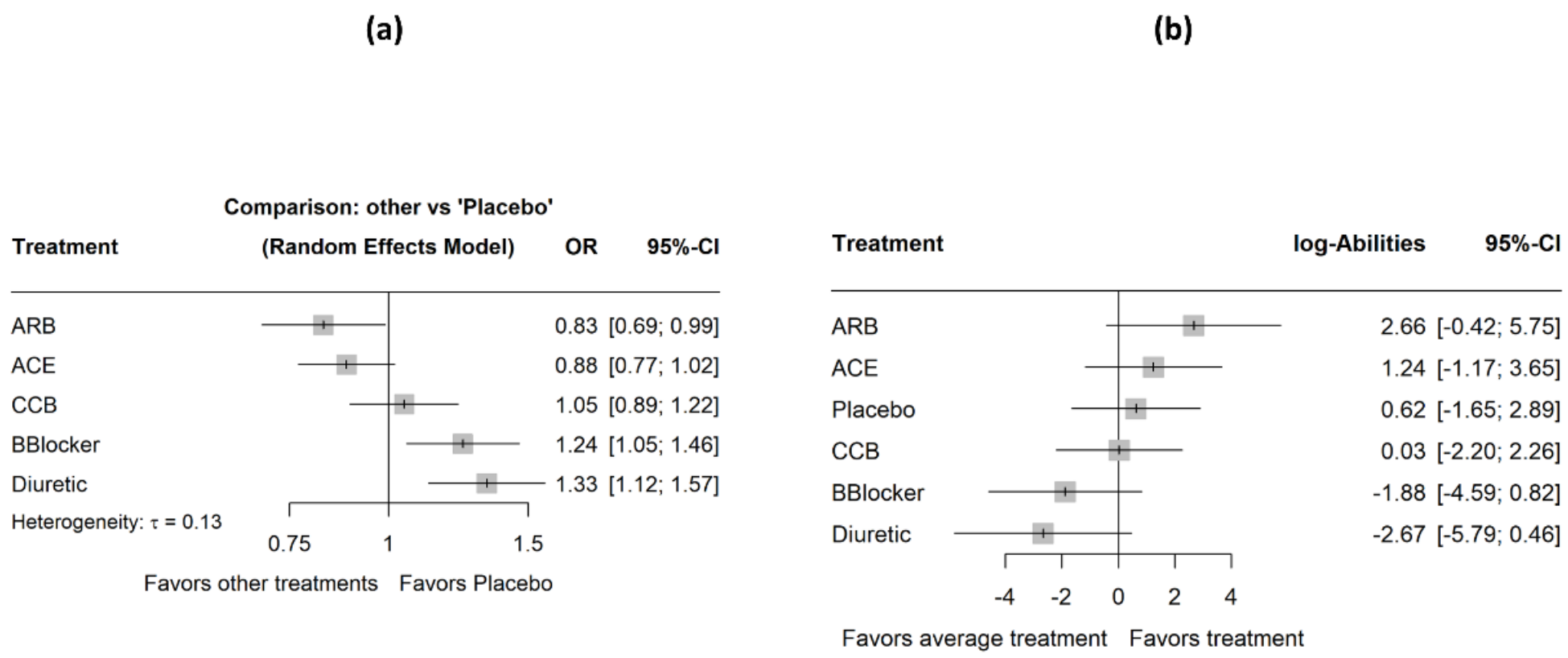

**Figure 5:** Forest plots with results for the network of antihypertensive treatments. Panel (a) shows the summary odds ratios obtained assuming placebo as the reference treatment group. Panel (b) shows the ranking results obtained using the proposed methodology. Abbreviations, ACE: angiotensin-converting-enzyme inhibitors, ARB: angiotensin-receptor blockers, CCB: calcium-channel blocker, BBlocker: Beta Blocker

We consider again an SWD equal to 1.20[40] and the respective ROE ranging from 0.83 to 1.20. The ranking results obtained from the approaches (1) to (4) can be found in **Table 2**, along with the NMA estimates of all treatments versus placebo while the respective results in terms of median ranks are available in Supplementary Material, Table 2. In this network small treatment effect values indicate beneficial effects. The results in terms of the estimated treatment abilities are depicted in **Figure 5b** while the normalized ability estimates $\hat{\pi}_X$ are shown in **Table 2**.

**Table 2:** Ranking metrics for the network of antihypertensives. Treatments with the top three values for each respective metric are shown in bold. The "Treatment" column is ordered according to P-scores.

| **Treatment** | **Odds ratio of Treatment vs Placebo [position in ranking]** | **P-scores** | **P-scores (SWD)** | **PReTA** | $p_{BV}$ | $\hat{\pi}_X$ |
|---|---|---|---|---|---|---|
| ARB | **0.83 [1]** | **0.95** | **0.67** | **1.00** | **0.76** | **0.69** |
| ACE | **0.88 [2]** | **0.84** | **0.52** | **1.00** | **0.23** | **0.16** |
| Placebo | **1 [3]** | **0.55** | **0.32** | **0.77** | 0.01 | **0.09** |
| CCB | 1.05 [4] | 0.46 | 0.24 | 0.45 | 0.00 | 0.05 |
| BBlocker | 1.24 [5] | 0.16 | 0.02 | 0.00 | 0.00 | 0.01 |
| Diuretic | 1.33 [6] | 0.04 | 0.00 | 0.00 | 0.00 | 0.00 |

Based on the NMA estimates, ARB showed the most beneficial treatment effect, closely followed by ACE, which had a similar estimate in both magnitude and precision. Regarding the other ranking metrics, there is complete agreement across all five approaches, with ARB consistently ranked first. Notably, the TCC in this network indicated that 63% of all NMA estimates yielded a treatment preference. This perfect agreement among ranking methods can likely be attributed to the low uncertainty in the treatment effect estimates. Specifically, the standard errors of the NMA estimates range from 0.07 to 0.10 (Supplementary Material, Figure 3). Finally, to assess the robustness of the estimated rankings with respect to the definition of the TCC, we performed a sensitivity analysis, progressively increasing the SWD in 0.10 increments from the recommended value of 1.20 up to 1.50. The results are shown in **Figure 6**. Overall, this sensitivity analysis showed that ARB and ACE remained the top two treatments across the different SWD values.

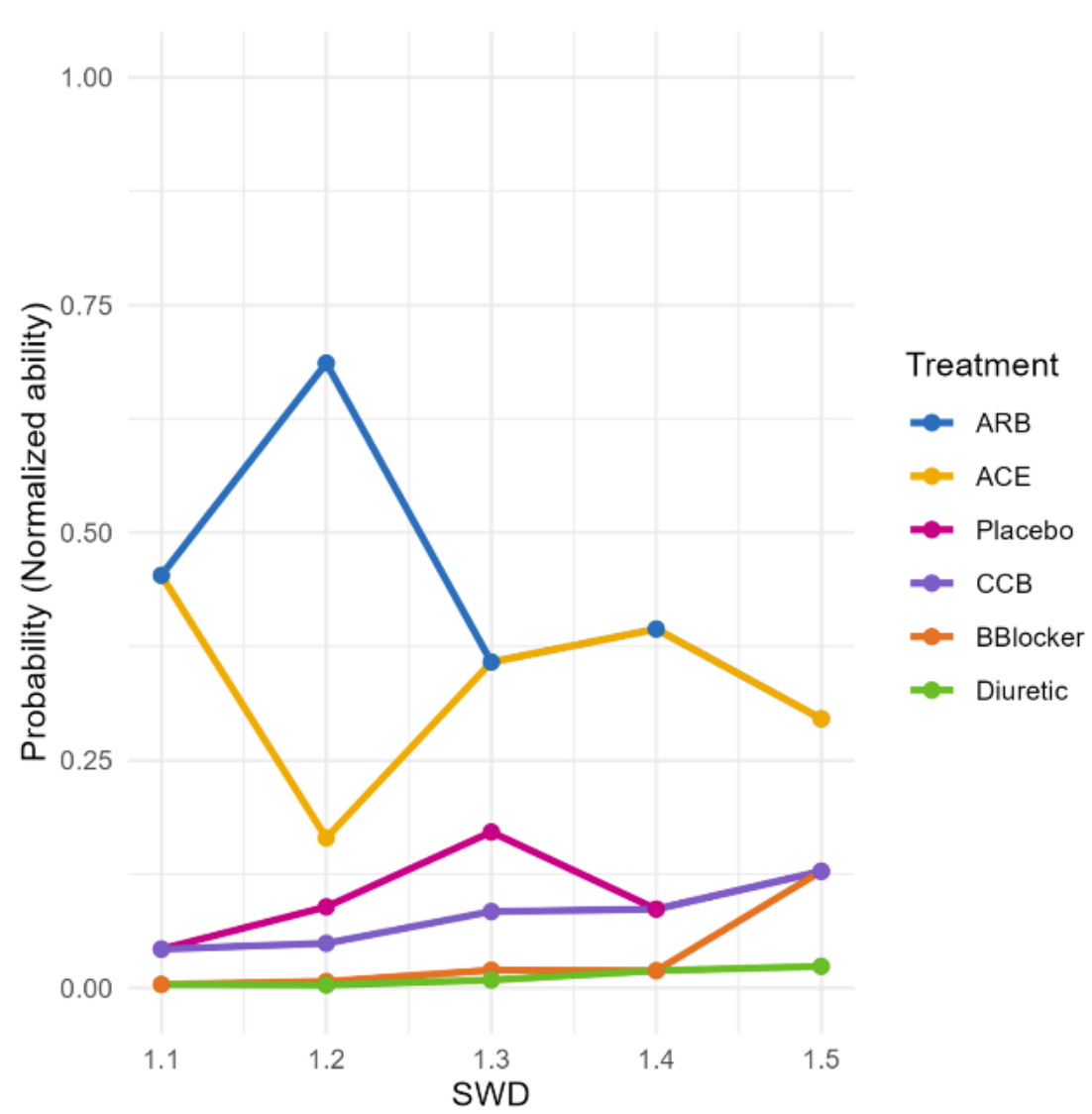

**Figure 6:** Sensitivity analysis for the network of the antihypertensive drugs. The y-axis represents the probability of each treatment having the highest true ability and the x-axis the different SWD values.

# 4 Empirical investigation across 153 published networks

## 4.1 Database

We studied the agreement across different ranking metrics by re-analysing networks from a database of published NMAs between 1999 and 2015, which included at least four treatments. To access these data we used the R package **nmadb**[41]. More details about this database can be found in the original publications[22,23]. In this database, 267 datasets were identified with available data. Given that there was no information regarding the SWD across these 267 networks, we used the recommendations from previous publications, which suggested that a common choice for the SWD in the case of the risk ratio (RR) would be a value of 1.25[42,43]. We therefore further restricted the database to include only networks with a binary outcome of interest. This yielded a database of 186 networks. After re-analysing these 186 networks, we obtained results from 174 networks, as 12 networks from **nmadb**[41] had incompatible data that did allow to fit a NMA model. Finally, applying the proposed ranking method to the set of 174 networks further restricted the networks with results to 153, as in the remaining 21 networks only ties were identified by the TCC. The NMA estimates and network geometries of these 21 networks are available in Supplementary Material's Figures 3-23.

## 4.2 Evaluated methods and performance metrics

We evaluated the agreement of the five methods presented in Section 3 in the context of a random-effects NMA model. This resulted in a total of 10 pairwise agreement comparisons between the different ranking metrics. Agreement was measured using Pearson's correlation coefficient, indicating the agreement in the ranking values obtained by each of the different ranking metrics. In other words, we investigated whether larger values in one ranking metric also corresponded to larger values in the other ranking metrics. This approach slightly deviates from previous works[5,7], which studied agreement between different ranking methods by investigating the agreement in the treatment order of the ranking list. This was not straightforward in our case, as the five methods of interest present the final treatment order in different ways (i.e., allowing for tied positions or always yielding an explicit order). Finally, we further investigated how the precision of the NMA estimates, as a measure of the total amount of information in the network, impacts the agreement between the proposed ability-based metric and the other ranking metrics. To this end, following Chioccia et al.[5], we contrasted the correlation coefficients from each of the 153 networks with the following measures:

i. the average variance across the $\binom{T}{2}$ NMA estimates $\widehat{\boldsymbol{\theta}}$,

ii. the relative range of variances, defined as $\frac{\max\{var(\widehat{\boldsymbol{\theta}})\}-\min\{var(\widehat{\boldsymbol{\theta}})\}}{\max\{var(\widehat{\boldsymbol{\theta}})\}}$.

## 4.3 Results

The results regarding the median correlation and the interquartile range (IQR) of correlations across the 153 networks are presented in **Table 3.** Overall, the proposed ability-based ranking metric was found to be strongly correlated with most other ranking metrics, as the median correlation coefficient was typically above 0.90. A similarly high level of agreement was observed among most of the alternative ranking methods. It is worth noting that the strong agreement between the P-score, P-score (SWD), and PReTA metrics was expected, given that P-score (SWD) and PReTA are essentially variations of the standard P-score approach. Finally, the agreement between $p_{BV}$ and the proposed method was generally moderate. The latter also applies in terms of the agreement between $p_{BV}$ and the rest of the evaluated method, with the correlation becoming stronger primarily when $p_{BV}$ was compared to the P-scores adjusted for SWD.

**Table 3:** Pairwise agreement between the different ranking metrics, measured by the median Pearson's correlation coefficient and the interquartile range of values obtained across 153 published NMAs.

| | **Median correlation [IQR]** |
|---|---|
| Ability based metric vs P-score | 0.949 [0.914, 0.973] |
| Ability based metric vs $p_{BV}$ | 0.653 [0.522, 0.798] |
| Ability based metric vs PReTA | 0.926 [0.868, 0.963] |
| Ability based metric vs P-score (SWD) | 0.931 [0.882, 0.974] |
| P-score vs $p_{BV}$ | 0.784 [0.709, 0.846] |
| P-score vs PReTA | 0.973 [0.958, 0.983] |
| P-score vs P-score (SWD) | 0.990 [0.977, 0.996] |
| $p_{BV}$ vs PReTA | 0.701 [0.589, 0.811] |
| $p_{BV}$ vs P-score (SWD) | 0.858 [0.774, 0.904] |
| PReTA vs P-score (SWD) | 0.963 [0.936, 0.982] |

**Figure 7** shows the results regarding the impact of uncertainty in the NMA estimates on the agreement between the ability-based metric and the other ranking metrics. In Panel (a), the different correlation coefficients were plotted against the average variance of the NMA estimates, which were log-transformed to enhance visibility. The overall trend suggests that as the average variance of the NMA estimates increases, the correlation between the ability-based metric and the other ranking metrics decreases. In Panel (b), the correlation coefficients were plotted against the relative range of variances. Following previous studies[5], the x-axis values were transformed using the double logarithm of the inverse relative range, so that values on the left-hand side indicate a larger variance range. These results showed that as the range of variances across the NMA estimates decreases, the agreement between the ability-based metric and the other ranking approaches increases. In other words, greater agreement is achieved in networks where the NMA treatment effects are estimated with similar levels of uncertainty. This is in line with previous empirical results that evaluated the rest of the approaches in terms of the same metrics[5,7].

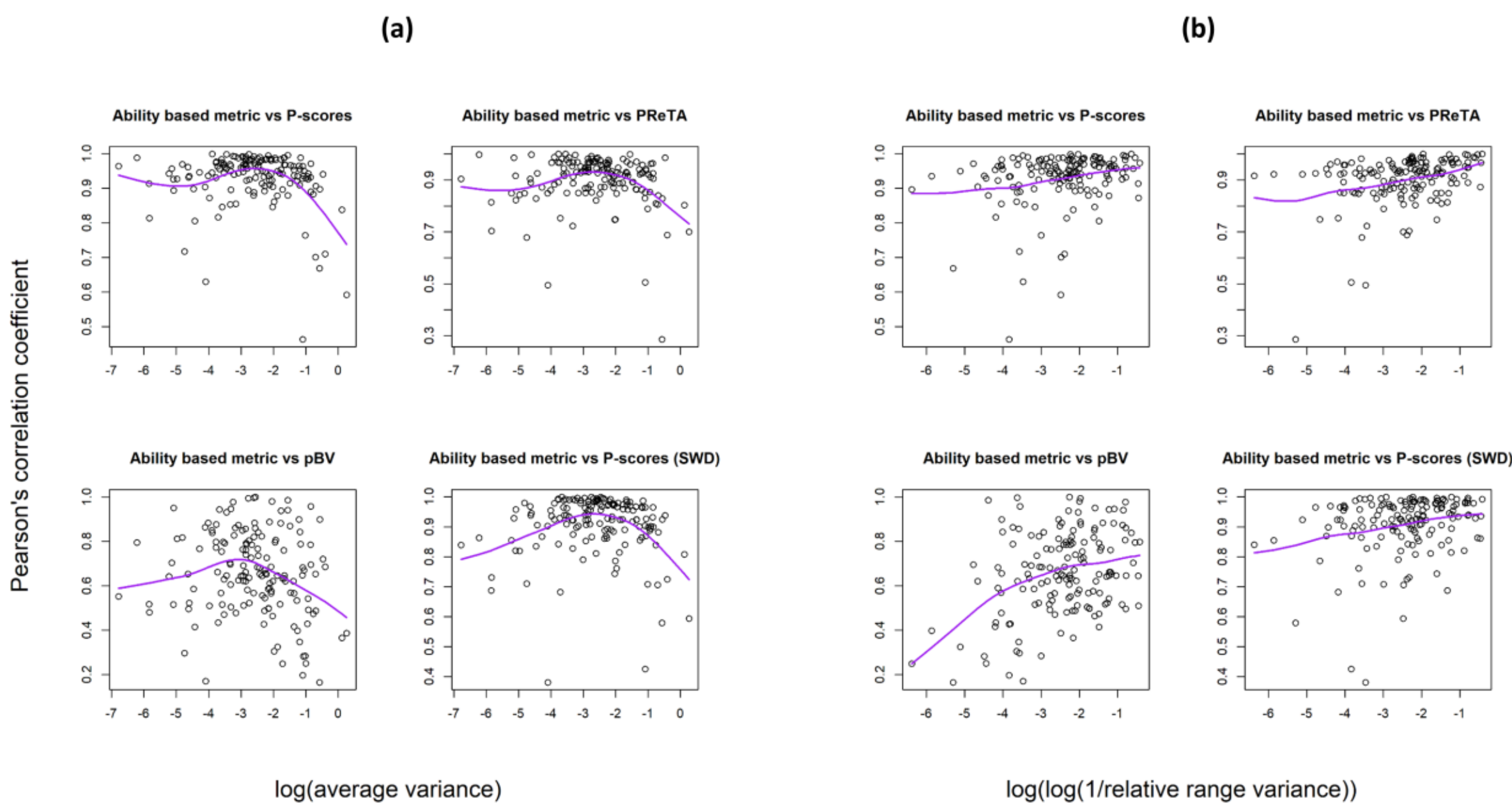

**Figure 7:** Scatter plots contrasting the correlation between the ability-based metric and the other ranking metrics across 153 networks. Panel (a) shows the correlations plotted against the average variance of the NMA estimates; values on the left-hand side of the graph indicate greater precision. Panel (b) shows the correlations plotted against the relative range of variances of the NMA estimates; values on the left-hand side of the graph indicate a larger variance range. In all scatter plots, the purple line represents a cubic smoothing spline with five degrees of freedom.

# 5 Discussion

In this paper, we introduce a novel framework for producing treatment hierarchies in NMA through a probabilistic ranking model that accounts for a pre-defined TCC. The rationale behind the proposed ranking method differs from existing approaches, as it combines the NMA estimates with a concrete TCC or, in other words, a decision rule—into a treatment hierarchy, whereas existing methods translate NMA estimates directly into rankings.

Our approach follows the principles of a typical decision-making process where a concrete decision rule is applied to the available evidence to translate the numerical results into practice[24,44–47]. We start by applying the predefined TCC to the NMA relative treatment effects, transforming them into treatment preference data. We propose as a clinically relevant TCC the ROE between two treatments that represents the area within which their relative effect lacks indication of a treatment preference[24,40]. Following previous work, we define the ROE using the SWD and its reciprocal (or opposite) value[24,40]. Here, we propose a simple way for defining an ROE-based TCC based on the magnitude of the NMA treatment effect and its uncertainty. However, any TCC considered appropriate and clinically relevant can be used by investigators to produce preference data.

We parameterize our model to estimate the ability of each treatment to outperform the other treatments in the network[28,29,31]; that is a latent characteristic referring to the propensity of each treatment in the network to yield clinically important and beneficial true treatment effects in the context of the defined TCC. Consequently, treatments with larger ability estimates corresponding to higher positions in the final ranking. Confidence intervals can also be placed next to the ability estimates to representing the uncertainty around the ranking metric. This should not be confused with other metrics proposed to evaluate the uncertainty of the treatment hierarchy[48]. Furthermore, the interpretation of the ability estimates also stems from their transformation into probabilities using Luce's axiom of choice[32,33]. Model diagnostics were recently developed and can also be investigated in cases where no ties are identified from the TCC[49]. However, these have not yet expanded to allow for ties. Overall, our method aims to produce clinically relevant treatment hierarchies accompanied by uncertainty measures. Of course, the proposed ranking method, like all existing ranking metrics, is not a substitute of the NMA relative effects; instead, it can be used to assist decision-making and treatment recommendations.

Establishing a direct one-to-one mathematical relationship between the true ability of each treatment and the true treatment effects is challenging, as the former is a latent characteristic dependent on the TCC, while the latter is a fixed unknown parameter. This complicates the design of simulation studies, which typically begin by defining true treatment effects. However, this challenge is not unique to our method but applies broadly to treatment ranking in NMA, as the scope of the existing ranking metrics is to summarize evidence based on NMA estimates and they cannot be calculated directly from true treatment effect values.

We used two published networks to assess the properties of our method and compare it with existing approaches. The network of antidepressants[20] represents an extreme case as the treatment ranked highest in terms of effect size (Vortioxetine) yielded the least precise NMA estimates. Using a TCC defined according to the SWD reported in the original publication[20], our method produced more conservative results than the other methods, particularly regarding Vortioxetine's position in the ranking. In a sensitivity analysis where we progressively increased the SWD, Vortioxetine moved to the top of the treatment hierarchy, reflecting its larger NMA estimate relative to other treatments. In the second network of antihypertensive treatments[21] we found a perfect agreement in the final ranking across all approaches. This agreement can be partly attributed to the high precision and narrow variance range of the NMA estimates.

We further explored the performance of the proposed framework and other common ranking metrics through a re-analysis of 153 published networks obtained from a published database[22,23], accessed via the R package **nmadb**[41]. This empirical study showed strong agreement among most of the evaluated ranking metrics, except for $pBV$, which exhibited only moderate agreement with the others. We also investigated how the total amount of information in a network, expressed as the uncertainty in NMA estimates, affects the agreement between the proposed ability-based metric and the other methods. The results indicated that agreement depends on the level of uncertainty: greater agreement was observed in networks where NMA estimates had higher precision and similar levels of uncertainty across treatments.

We see several advantages of our proposed treatment ranking approach. First, the requirement of a priori defining a concrete TCC enables researchers to consider early on what constitutes a preferred treatment. In our approach we estimate the treatment ability using maximum likelihood theory, thereby allowing us to obtain the standard error of the estimated abilities and infer about the uncertainty of ranking positions using standard statistical measures. In addition, the proposed model does not provide treatment ability estimates when all the NMA treatment effect estimates indicate ties due to convergence failure. Although this might be considered as a drawback of the model, we see it also as a way of preventing researchers from making ranking statements in the absence of sufficient evidence that the NMA estimates fulfil the TCC. This is in line with previous NMA recommendations for avoiding the presentation of ranking results in the presence of large uncertainty in the relative effects[5].

Despite these advantages, our approach is not free of limitations. Probably the most important limitation relates to the definition of the SWD and of the respective ROE that involves some subjectivity[26]. On the other hand, though, the use of different ROEs allows researchers to estimate the treatment hierarchy under different settings (e.g. for different patient profiles). Ways to mitigate this inherent subjectivity have been suggested in the literature through fully statistical approaches[25] or by incorporating information from patients[25]. Moreover, investigators conducting NMAs may choose to define another TCC not based on the ROE. To avoid data-driven decisions, we recommend meta-analysts using our ranking method to define and justify the TCC they plan to use in their protocol and investigate the robustness of the estimated hierarchy under different SWD values.

Our proposed framework offers a novel alternative to existing ranking metrics for estimating treatment hierarchies in NMA. The importance of a well-defined treatment hierarchy question prior to estimating treatment ranking has been highlighted recently[6]. To our knowledge, this is the first approach that incorporates explicitly and quantitatively considerations on the treatment hierarchy question through the pre-defined TCC. Future extensions of the proposed approach could include adapting the model to account for treatment-level characteristics (e.g. treatment cost) and multiple outcomes. The former is currently possible only in cases where no ties are allowed from the TCC[50]. Overall, investigators can use the proposed approach either as their primary ranking tool or as sensitivity analysis alongside conventional ranking metrics particularly for networks with increased uncertainty in their relative effects and knowledge of clinically relevant TCC.

**Author contributions:** Theodoros Evrenoglou: conceptualisation; analysis; software; writing - original draft; writing – review and editing. Adriani Nikolakopoulou: writing – review and editing. Guido Schwarzer: software; analysis; writing – review and editing. Gerta Rücker: writing – review and editing. Anna Chaimani: conceptualisation; writing - original draft; writing – review and editing.

**Acknowledgments:** We would like to thank Prof. Georgia Salanti and the two anonymous reviewers for their valuable comments on earlier versions of the manuscript.

**Funding:** Theodoros Evrenoglou received funding by the French National Research Agency under the project ANR-22-CE36-0013-01 and from the Deutsche Forschungsgemeinschaft (DFG, German Research Foundation) under the Project ID-554095932. Adriani Nikolakopoulou was supported by DFG - grant number NI 2226/1-1 and Project-ID 499552394 – SFB 1597.

**Conflicts of interest statement:** None.

**Data availability statement:** The full code and data to reproduce the results of the two illustrative examples and the empirical study is freely available on GitHub using the following link: https://github.com/TEvrenoglou/ranking_article.